\documentclass{report}
\usepackage{svcon2e}
\usepackage{amssymb}
\usepackage{epsf}

 \newcommand{\bm}[1]{\mbox{\boldmath $#1$}}
 \newcommand{\bms}[1]{\mbox{\footnotesize \boldmath $#1$}}
 
 \newcommand{\zr}[1]{\mbox{\hspace*{#1em}}}

 \newcommand{\Id}{\mbox{1\zr{-0.62}{\small 1}}}

 \newcommand{\RR}{\mathbb R}
 \newcommand{\CC}{\mathbb C}
 \newcommand{\ZZ}{\mathbb Z}

 \newcommand{\QQ}{\mathbb Q}

 \newcommand{\h}{\mbox{\small $\#$}}

 \newcommand{\cf}{\mathbf 1}

\newtheorem{defin}{Definition}

\newtheorem{prop}{Proposition}

 \newcommand{\be}{\begin{equation}}
 \newcommand{\ee}{\end{equation}}

\begin{document}
\pagenumbering{arabic} 
%\tableofcontents 

\chapter{Self-similarities and invariant densities for model sets}
\chapterauthors{Michael Baake{}\footnote{Institut 
f\"ur Theoretische Physik, 
Universit\"at T\"ubingen, Auf der Morgenstelle 14,
D-72076 T\"ubingen, Germany}
and Robert V. Moody{}\footnote{Department of Mathematical
Sciences, University of Alberta, Edmonton, Canada, T6G 2G1}}

\begin{abstract} 
Model sets (also called cut and project sets) are generalizations of 
lattices. Here we
show how the self-similarities of model sets are a natural
replacement for the group of translations of a lattice. This leads
us to the concept of averaging operators and invariant densities
on model sets. We prove that invariant densities exist and that they
produce absolutely continuous invariant measures in internal space. 
We study the invariant densities and their relationships
to diffraction, continuous refinement operators, and Hutchinson 
measures.
\end{abstract}

\section{Model sets and self-similarities} 
\label{starting}
In this paper we introduce the notion of averaging operators
on suitable spaces of functions on model sets. An averaging operator
encodes information about the entire set of self-similarities with
given inflation factor for a given model set. It can be 
interpreted as a Hilbert-Schmidt operator on  
the space of continuous functions on the corresponding acceptance
window and, remarkably, from this point  of view 
is seen to be an example of the recently studied continuous
refinement operators. Using this connection we can determine
the spectrum and associated set of eigenfunctions for any 
inflation factor
of any given model set. In particular,
the leading eigenvalue $1$ gives rise to an {\em invariant density\/}
for the model set. We derive some properties of the Bragg
spectrum of a model set that has been weighted by an invariant 
density. 
We also show that an invariant density leads to an absolutely
continuous invariant measure
on internal space and we relate this measure to a weakly 
converging sequence of Hutchinson measures. The 
full mathematical development of this work will appear in \cite{BM}. 

\subsection{Cut and project schemes}
We begin with the notion of a cut and project scheme. By definition,
this consists of a collection of spaces and mappings: 

\be \label{cutandproject}
  \begin{array}{ccccc}
   \RR^m & \stackrel{\pi^{}_1}{\longleftarrow} & \RR^m \times \RR^n &
           \stackrel{\pi^{}_2}{\longrightarrow} & \RR^n  \\
    & & \cup & & \\ & & \tilde{L} & & \end{array}
\ee
where $\RR^m$ and $\RR^n$ are two real spaces, $\pi^{}_1$ and 
$\pi^{}_2$
are the projection maps onto them, and $\tilde{L} \subset 
\RR^m \times \RR^n$
is a lattice. We assume that $\pi^{}_1|^{}_{\tilde{L}}$ is
 injective and
that $\pi^{}_2(\tilde{L})$ is dense in $\RR^n$. We call $\RR^m$
(resp.\ $\RR^n$) the physical (resp.\ internal) space. 
We will assume that
$\RR^m$ and $\RR^n$ are equipped with Euclidean metrics and that
$\RR^m \times \RR^n$ is the orthogonal sum of the two spaces. For $x$
lying in any of these spaces, $|x|$ denotes its length.

A cut and project scheme involves, then, the projection of
a lattice into a space of smaller dimension, but a lattice that is 
transversally located with respect to the projection maps involved.

\subsection{Example}\label{ex1}

A simple, and very useful, example of such a scheme arises from
a real quadratic irrationality $q$. We form the ring 
$\ZZ[q] \subset \RR$
and let $^*$ be the $\ZZ$-mapping that takes $q$ into
its algebraic (quadratic) conjugate. Then the set of points 
$\widetilde{\ZZ[q]} := \{(x,x^*)~ | ~ x \in \ZZ[q]\}$
is a lattice in $\RR^2$ and 

\be \label{qcutandproject}
  \begin{array}{ccccc}
   \RR & \stackrel{\pi^{}_1}{\longleftarrow} & \RR^2 &
           \stackrel{\pi^{}_2}{\longrightarrow} & \RR  \\
    & & \cup & & \\ & &\widetilde{\ZZ[q]} & & \end{array}
\ee
where we use the coordinate projections, is a cut a project scheme.
An important case of this occurs when $q=\tau := (1+\sqrt{5}\,)/2$.

\subsection{Model sets} \label{sec-1.3}
Let $L := \pi^{}_1(\tilde{L})$ and let
\be \label{star}
     (\,)^* \, : \quad L \; \longrightarrow \; \RR^n
\ee
be the mapping $\pi^{}_2 \circ (\pi^{}_1|^{}_{\tilde{L}})^{-1}$.
This mapping extends naturally to a mapping on the rational span 
$\QQ L$ of $L$, also denoted by $(\,)^*$.  Note that the lattice 
$\tilde{L}$ can also be written as
\be
    \tilde{L} \; = \; \{ (x,x^*) \mid x \in L \}.            
\ee
Now, let $\Omega \subset \RR^n$. Define
\be
     \Lambda \; = \; \Lambda(\Omega) \; := \; 
                     \{ x \in L \mid x^* \in \Omega \, \} \, .
\ee
We call such a set $\Lambda$ a {\em model set} 
(or {\em cut and project set}) 
if the following three conditions are fulfilled,
\begin{itemize}
\item[\bf W1] $\Omega \subset \RR^n$ is compact.
\item[\bf W2] $\Omega \;=\; \overline{\mbox{int}(\Omega)}$.
\item[\bf W3] The boundary of $\Omega$ has Lebesgue measure 0.
\end{itemize}

The mathematical reasons for studying model sets are
that they are very natural generalizations of lattices, they
share many properties with them, and they
allow symmetries that are otherwise unavailable 
in lattices of the corresponding dimensions.
For example, the following properties are shared by all
model sets $\Lambda$:
\begin{itemize}
\item[\bf M1] $\Lambda$ is {\em uniformly discrete}: that is to say,
there is an $r>0$ so that for all distinct $x,y \in \Lambda$, ~
$|x-y| \ge r$.
\item[\bf M2] $\Lambda$ is {\em relatively dense}: that is to say,
there is an $R>0$ so that for each $x \in \RR^m$ the open ball of
radius $R$ around $x$ contains a point of $\Lambda$. 
\item[\bf M3] There is a finite set $F$ so that 
$\Lambda - \Lambda \subset \Lambda - F$.
\item[\bf M4] $\Lambda - \Lambda$ is a Delone set (i.e. satisfies
[{\bf M1, M2}]).
\item[\bf M5] $\Lambda$ has a well-defined density $d$, i.e.\
\be \label{1.4}
     d \; = \; \lim_{s\rightarrow\infty}
        \frac{\h \Lambda_s}{\mbox{vol}(B_s(0))}
       \; = \; \lim_{s\rightarrow\infty}
        \frac{\h \Lambda_s}{c_m s^m}
\ee
exists, where $B_s(0) := \{ x \in \RR^m \mid |x| \leq s \}$ and
\be \label{unitsphere}
     c_m \; := \; \frac{\pi^{m/2}}{\Gamma({m\over 2} + 1)}
\ee
is the volume of the unit sphere in $\RR^m$.
\item[\bf M6] $\Lambda$ diffracts. (See section 3.1 for
more on this.)
\end{itemize}

A set with the properties {\bf M1} and {\bf M2} is called a 
{\em Delone} set. A lattice is nothing else than a Delone set
that is a group. If $F = \{0\}$ then {\bf M3} states that $\Lambda$ 
is a group, so {\bf M3} is in fact a generalization of the
group law. The limit in (\ref{1.4}) is easily seen to be independent 
of the choice 
of origin for the Euclidean space. What is more, it even exists
{\em uniformly} for sets. This means that for any subset $S$ of 
$\Omega$ with boundary of measure $0$, the relative frequency
of the points of $(\Lambda_s)^*$ falling into $S$, as $s \to \infty$, 
is ${\rm vol}(S)/{\rm vol}(\Omega)$, and the convergence is uniform
with respect to translation of the set $S$.
For more on these properties one may consult \cite{ Hof,Meyer,
Moody, Martin2}.

Model sets arise in situations in which one is looking for
Delone structures with symmetries that are incompatible with
lattices. The most famous example is that of the icosahedral group
which cannot appear as the point symmetry of any lattice
in $3$-space. It is known \cite{Peter, jlvg} that if 
$G$ is any finite group acting
irreducibly in $\RR^m$ and $X$ is any non-trivial orbit of $G$,
then either $G$ acts crystallographically in $\RR^m$, that is
to say, there is a $G$-stable lattice of $\RR^m$; or
there is a $G$-stable cut and project set in $\RR^m$
that contains the set $X$. This is the origin of the interest
in these sets in the theory of quasicrystals, see \cite{Janot}
for background material.

However, there is a serious price to be paid for moving
from lattices to model sets. Lattices, by definition,
have an entire lattice of translational symmetries. By
comparison, in a model set $\Lambda$ described
by a cut and project scheme (\ref{cutandproject}), 
the set of translational
symmetries is the kernel of $()^*$ in (\ref{star}),
 and in all the standard examples
this is in fact $\{0\}$. Fortunately, in many cases of interest,
there is nonetheless an abundance of symmetry, as long as one 
is prepared to consider self-similarities instead of group symmetries.

\subsection{Self-similarities}
\begin{defin} 
A {\em self-similarity}
of $\Lambda$ is an affine linear mapping $t=t^{}_{Q,v}$
\be \label{selfsim}
     t^{}_{Q,v} \; : \quad x \mapsto Qx + v
\ee
on $\RR^m$ that maps $\Lambda$ into itself, where $Q$ is a (linear) 
similarity
and $v\in\RR^m$. Thus $Q = q R$, i.e.\ it is made up of an orthogonal 
transformation $R$ and an {\em inflation factor} $q$.
\end{defin}

Let $t^{}_{Q,v}$ be a self-similarity of $\Lambda$. If $\Lambda$ is
uniformly discrete, we must have $|q|\geq 1$. This is the reason
why we also call such a self-similarity an {\em affine inflation}. 
We are interested in the {\em entire} set
of affine inflations with the same similarity factor $Q$. 
It is convenient
to have $0\in\Lambda$ and $0 = 0^* \in {\rm int}(\Omega)$.
Using a translation of $\Lambda$ by a vector
$v^{}_0\in L$ and the corresponding translation of $\Omega$ by
$v^*_0\in L^*$, we assume this to be the case. This makes no 
structural difference to the set of inflations of $\Lambda$, but
simplifies the algebra: if $0\in\Lambda$ and $t^{}_{Q,v}$ is an affine
inflation, then $v\in L$ and $Q(L)\subset L$. In fact, we are going to
also assume that $QL=L$.

Let us then fix once and for all a (linear) similarity transformation
$Q$ on $\RR^m$ such that $QL = L$. How do we describe the
set of all self-similarities with similarity factor $Q$? 
In preparation
for answering this question, it is useful to note that 
there are three different ways of looking
at the same cut and project set $\Lambda$: first as a Delone 
set in $\RR^m$,
which we may think of as the {\em discrete} picture; second
as part of the lattice $\tilde{L}$, which we may think of 
as the {\em arithmetic} picture; and finally as a dense subset of
$\Omega$ via the mapping $()^*$, which may be thought
of as the {\em analytic} picture. 

As an illustration of these ideas,
note that $Q$ naturally gives rise to an automorphism $\tilde{Q}$ of
the lattice $\tilde{L}$, i.e. an element of $GL_{\ZZ}(\tilde{L})$,
and a linear mapping $Q^*$ of $\RR^n$ that maps
$\Omega$ into itself. From the arithmetic nature of $\tilde{Q}$
we deduce that the eigenvalues of $Q$ and $Q^*$ are algebraic
integers and from the compactness of $\Omega$ that 
$Q^*$ is contractive. Furthermore, one can deduce \cite{BM} 
that $Q^*$ is
diagonalizable from the corresponding property of $Q$. Strictly
speaking we can only deduce that the eigenvalues of $Q^*$ 
do not exceed
$1$ in absolute value, but we will always assume that in fact
they are less than one in absolute value. In the sequel we will
normally denote the contraction $Q^*$ by $A$ to match various
sources we will refer to frequently.

Define 
\be \label{2.8}
      \Omega^{}_Q \; := \; \{ u \in \RR^n \mid
                A \Omega + u \subset \Omega \} \, ,
\ee
we say that $Q$ is {\em compatible} with $\Lambda$ if
$\mbox{int}(\Omega^{}_Q)\neq\emptyset$. In this article, we shall
always assume that not only $\Omega$, but also $\Omega^{}_Q$ is
Riemann measurable, i.e. $\partial \Omega^{}_Q$ has zero Lebesgue
measure. Interpreting
(\ref{selfsim}) on the window side we obtain:
\begin{prop}
Let $\Lambda = \Lambda(\Omega)$ be a model set based on a window
$\Omega$ that satisfies the window conditions {\bf W1} -- {\bf W3}.
Let $Q$ be a similarity compatible with $\Lambda$. 
Then the set ${\cal T}_Q$ of
affine inflations with the same similarity $Q$ is the set of mappings
$t^{}_{Q,v}$ where $v$ runs through the set
\be \label{2.9}
     T^{}_{Q} \; = \; \{ v\in L \mid v^* \in \Omega^{}_Q \} \, .
\ee
In particular, $T^{}_{Q}$ is also a model set.
\end{prop}

Let us pause to consider the special situation where $\Omega$ is 
convex. 
 In this case, $\Omega^{}_Q$ is also convex
and hence satisfies the conditions that we need. If in addition
$Q^* = \varepsilon \!\cdot \Id\,$,\, $0<\varepsilon<1$, 
which actually is often the case in examples of physical relevance, 
one obtains
\be \label{eps-omega}
     \Omega^{}_Q \; = \; (1 - \varepsilon) \Omega \, .
\ee 
If $-1 < \varepsilon < 0$, but $\Omega = - \Omega$, (\ref{eps-omega})
is still true if $\varepsilon$ is replaced by $|\varepsilon|$.
This happens in our Examples.

\subsection{Example} \label{ex2}
In Example \ref{ex1} above, take $q = \tau$.
A simple model set is defined by
\be
    \Lambda \; = \; \{ x \in \ZZ[\tau] \mid x' \in [-1,1] \}.
\ee
Let us look at the inflation factor $q=\tau$. 
Multiplication by $q$ determines
the contraction $A=\tau' \!\cdot \Id\,$, $\tau'=-1/\tau$, 
on the internal side and
\be
   \Omega^{}_{\tau} \; = \; 
   \{ u\in\RR\mid -\frac{1}{\tau}\cdot [-1,1] + u \subset [-1,1] \}
   \; = \; [-1 +\frac{1}{\tau} \, , \, 1 - \frac{1}{\tau}] \, .
\ee
Thus, for all $v\in\ZZ[\tau]$ with $v'\in [-1/\tau^2 , 1/\tau^2]$,
\be
     t^{}_{\tau,v} \; : \quad x \mapsto \tau x + v
\ee
is an inflation of $\Lambda$, and
\be
     t^*_{\tau,v} \; : \quad  y \mapsto \frac{-y}{\tau} + v'
\ee
is the corresponding contraction in internal space.

\section{Averaging operators and invariant densities}

One of the most famliar techniques in the theory of group
representations is the use of group averages in order to produce
invariants.  Thus for a finite group $G$ one typically invokes
the process 
\be
    F \; \mapsto \; \frac{1}{\h G} \sum_{g\in G} g\cdot F
\ee
which averages the function $F$ over the group, where
$g\cdot F (x) := F(g^{-1} x)$. We intend to 
do exactly the same thing replacing $G$ by the set of all
self-similarities ${\cal T}_Q$ of a model set $\Lambda$. Since 
${\cal T}_Q$ will be infinite we have to be a little
careful in averaging.

For any subset $T\subset\RR^m$ and for any $s\geq0$,
we thus define
\be \label{1.3}
      T_s \; := \; \{ x \in T \mid |x| \leq s \}
\ee
where $|x|=(x\cdot x)^{1/2}$ is the standard Euclidean norm on 
$\RR^m$.

\subsection{Averaging operators}
\begin{defin}  Let $\Lambda = \Lambda(\Omega)$ be a model set 
  based on a window
  $\Omega$ that satisfies the window conditions {\bf W1} -- {\bf W3}.
  Let $Q$ be a similarity (with inflation factor $q$) that is 
  compatible with $\Lambda$, let
  ${\cal T} = {\cal T}_Q$ be the set of all self-similarities of
  $\Lambda$ with
  similarity $Q$, and let $T = T_Q$ be the corresponding set of
  translations. Let $p: \, L \rightarrow \RR$ be any function
  on $L$ that vanishes off $\Lambda$. 
  Then the average of $p$ over $\cal T$
  is defined to be 
\be \label{2.12}
   ({\cal A}p)(x) \; = \; \lim_{s\rightarrow\infty} 
                          \frac{{|\rm det}(Q)|}{\h T_s}
                \sum_{v\in T_s} p(t_v^{-1} x)
\ee 
provided this limit exists. We say that $p$ is a $Q$-invariant
density on $\Lambda$ if
\begin{itemize}
\item[\bf ID1]  $p$ is non-negative on $\Lambda$,
\item[\bf ID2]  ${\cal A}p = p$, and
\item[\bf ID3]  $p$ is normalized, i.e.
\be \label{2.13}
    \lim_{s\rightarrow\infty}\, \frac{1}{\h \Lambda_s} 
          \sum_{x\in\Lambda_s} p(x) \; = \; 1 \, .
\ee
\end{itemize}
\end{defin} 

Note that for $x \in \Lambda$ and $v \in T$, $t_v^{-1} x \in L$ but 
$t_v^{-1}x$ does not in
general lie in $\Lambda$. Thus in (\ref{2.12}) one can expect that 
many of the summands on the right hand side will be $0$ because
$p$ vanishes off $\Lambda$.
Note also that (\ref{2.13}) is a normalization per point of $\Lambda$.
This can be changed to a normalization per unit volume if necessary
because the density $d$ of points of $\Lambda$ exists. 

Let ${\cal C}(\Omega)$ be the space of all continuous complex-valued 
functions on $\RR^n$ with support in $\Omega$. Via
the mapping $()^*$ of (\ref{star}) we obtain a space 
${\cal C}(\Lambda)$
of functions on $L$, vanishing off $\Lambda$: for 
$f \in {\cal C}(\Omega)$, 
we define $p \;=\; p^{}_f \in {\cal C}(\Lambda)$ by 
\be \label{3.2}
    p(x) \; = \; {\rm vol}(\Omega) \cdot f(x^*) \, ,
\ee
for $x \in \Lambda$, and $p(x)=0$ otherwise.
Here, the normalization constant ${\rm vol}(\Omega)$ is 
thrown in to make
things more convenient later on.

 Bearing in mind (\ref{3.2}), we rewrite (\ref{2.12}) as a new
averaging operator on ${\cal C}(\Omega)$:
\be
   ({\cal A}f)(x^*) \; = \; \lim_{s \rightarrow \infty} \, 
          \frac{|\det (Q)|}{\h T_s}
          \sum_{v\in T_s} f((t_v^*)^{-1} x^*)
\ee
for all $x^* \in \Lambda^*$. 

Now, it is well-known \cite{Martin,Martin2,Hof} that the points 
$\pi^{}_2(\tilde{L})$ are uniformly distributed in $\RR^n$, 
in the sense described in Section \ref{sec-1.3}.
In particular, the points of $T^*$ are uniformly distributed in
$\Omega^{}_Q$, hence $\lim_{s\rightarrow\infty} \overline{(T_s)^*}
= \Omega^{}_Q$. Using Weyl's theorem \cite{Weyl, Kuipers},
the continuity of $f$,  and the fact that 
$\Omega^{}_Q$ is Riemann integrable, we obtain
\be \label{3.5}
   ({\cal A} f)(x^*)\; = \; \frac{|\det(Q)|}{{\rm vol} (\Omega^{}_Q)}
                  \int_{\Omega^{}_Q} f(A^{-1}(x^* - u)) du \, .
\ee
Using the fact that $\det(Q) = \det(A)^{-1}$ (remember that
$A=Q^*$) and introducing
the normalized indicator (or characteristic) functions
\be \label{normChar}
      X_S \;:= {{\cf_S}\over {\rm{vol}(S)}}
\ee
defined for all measurable subsets $S$ of $\RR^n$, we can rewrite
(\ref{3.5}) in a number of equivalent ways:
\begin{eqnarray} \label{cntsRefin}
({\cal A} f)(x) & = & \int_{\RR^n} X_{Ay+\Omega_Q}(x) f(y) dy 
                      \nonumber\\
      & = & {1 \over {|\det(A)|}}\int_{\RR^n} X_{A^{-1}\Omega_Q} 
             (A^{-1}x -y) f(y) dy \nonumber \\
       & = & {1 \over {|\det(A)|}} 
             \int_{\RR^n}X_{\Omega_Q}(x-y) f(A^{-1} y) dy \,.
 \end{eqnarray}
This shows that the averaging operator ${\cal A}$ is a 
{\em continuous refinement operator} in the
sense of \cite{Jia} in the one-dimensional case and \cite{Jiang} in
the multi-dimensional case. 
In Fourier space, by application of the convolution theorem,
(\ref{cntsRefin}) reads as
\be \label{fourierCntsRef}
 \widehat{({\cal A}f})(k) = \widehat{X_{\Omega_Q}}(k)\hat{f}(A^tk) 
\ee 
where $\hat{f}(k) := \int_{\RR^n} e^{-2\pi i k\cdot x} f(x) dx$
and $f(x) = \int_{\RR^n} e^{2\pi i k\cdot x} \hat{f}(k) dk$.

\subsection{Invariant densities}
At this point, we can determine a function $f=f_p$ corresponding
to an {\em invariant density} $p$. 
We are looking for a $1$-eigenfunction
of the operator ${\cal A}$. The normalization condition {\bf ID3},
seen on the window side, becomes, using Weyl again,
$\int_{\RR^n} f(u) du = 1$.
Iteration of (\ref{fourierCntsRef}) leads to
\be \label{4.4}
    \hat{f}(k) \; = \; \hat{X}_{\Omega_Q}(k) \cdot 
                       \hat{X}_{\Omega_Q}(A^t k) 
\cdot \ldots \cdot
   \hat{X}_{\Omega_Q}((A^t)^N k) \cdot \hat{f}((A^t)^{N+1} k) \, .
\ee
Since $A$ is a contraction, so is $A^t$. Consequently, $(A^t)^{N+1} k
\rightarrow 0$ as $N\rightarrow\infty$. Since $\hat{f}$ and 
$\hat{X}_{\Omega^{}_Q}$ are
$C^{\infty}$ and $\hat{f}(0) = 1$, we can take the limit and obtain
\be \label{prodFormula}
   \hat{f}(k) \; = \;  \prod_{N=0}^{\infty}
              \hat X^{}_{\Omega^{}_Q} ((A^t)^N k)                  
               \; = \;  \prod_{N=0}^{\infty}
              \frac{\hat{\cf}^{}_{\Omega^{}_Q} ((A^t)^N k)}
                   {{\rm vol}(\Omega^{}_Q)}  \, .
\ee
This is an infinite product with compact convergence that
solves our problem: $\hat{f}$ is an infinite product of
$C^{\infty}$-functions, and is itself $C^{\infty}$.
The function $f$ is now the inverse Fourier transform
of $\hat{f}$. By construction, it is the Radon-Nikodym derivative,
and hence the density, of an absolutely continuous 
invariant (\ref{invariantdensity})
measure (with respect to Lebesgue measure) in internal space.  
Again applying the convolution theorem we arrive at:

\begin{prop} 
 Let $\Lambda = \Lambda(\Omega)$ be a model set based on a window
$\Omega$ that satisfies the window conditions {\bf W1} -- {\bf W3}.
Let $Q$ be a similarity compatible with $\Lambda$ and let 
$A:= Q^*$. Then there
is a unique $Q$-invariant density $p$ for
$\Lambda$ lying in ${\cal C}(\Lambda)$. 
This is given through $p=p^{}_f$, see {\rm (\ref{3.2})}, where $f$ is
given by the infinite convolution product
\be
    f \; = \; 
      \mathop{\mbox{\Huge $*$}}_{N=0}^{\infty}
      \frac{\cf^{}_{A^N \Omega^{}_Q}}
           {{\rm vol}(A^N \Omega^{}_Q)} \, .
\ee
\end{prop}
Note that this convolution of characteristic functions defines
a $C^{\infty}$ function with compact support contained in $\Omega$.

If $Q,Q^2,\dots, $ are all compatible with $\Lambda$ then it
is instructive to look at the corresponding invariant densities
$f = f_{(1)}, f_{(2)}, \dots $ The sets $\{\Omega_{Q^n}\}$
are increasing and $\overline{\bigcup_n \Omega_{Q^n}} = \Omega$.
The functions of the sequence $\{\hat{f}_{(k)}\}$ become increasingly 
concentrated around $0$ and it is natural to expect 
$\lim_{s \to \infty} f_{(s)} = \cf_\Omega/{\rm vol}(\Omega)$. 
This is illustrated in 
the case of our Example by the sequence of graphs of Fig.1.

\subsection{Example}

We continue with Example \ref{ex2} from above and determine
the invariant density (or rather its Fourier transform) on 
$[-1,1]$ corresponding to
the inflation factor $\tau$. If $f$ is this density,
then equation (\ref{fourierCntsRef}) becomes
\be
   \hat{f}(k) \; = \; \frac{\tau^2}{2} 
   \hat{\cf}^{}_{[-1/\tau^2,1/\tau^2]} (k) \hat{f}(-k/\tau) \, .
\ee
Routine calculation gives:
\be
  \frac{1}{2a} \hat{\cf}^{}_{[-a,a]}(k) 
  \; = \; \frac{1}{2a} \int_{-a}^{a} e^{-2\pi i kx} dx 
  \; = \; \frac{\sin(2\pi a k)}{2\pi a k} \, .
\ee
So, we obtain
\be
  \hat{f}(k) \; = \; \frac{\sin\left({2\pi k\over\tau^2}\right)}
             {\left({2\pi k\over\tau^2}\right)} \hat{f}(-k/\tau)
  \; = \; \ldots \; = \;
  \prod_{N=2}^{\infty} \frac{\sin\left({2\pi k\over\tau^N_{}}\right)}
              {\left({2\pi k\over\tau^N_{}}\right)} \, .
\ee
The calculations for other inflation factors are similar; Fig.~1
shows some of the resulting invariant densities.

\vspace*{8mm}
\begin{figure}[ht]
\centerline{\epsfysize=65mm \epsfbox{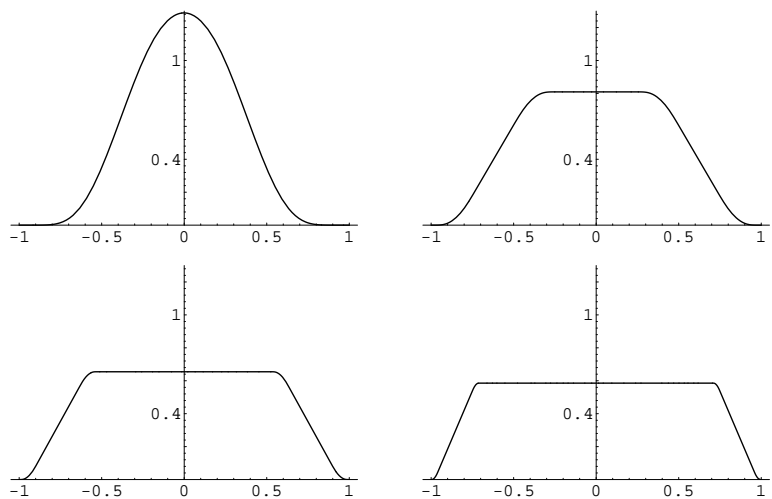}}
\caption{Invariant densities for the model set of Example \ref{ex2}, 
         for inflation factors $\tau$, $\tau^2$, $\tau^3$ and 
         $\tau^4$.}
\end{figure}

% ******************************************
% Figure 1 to go in here  4 invariant densities
% ******************************************

\subsection{Eigenvalues and eigenfunctions}
According to \cite{Jiang}, if the eigenvalues of $A^t$ are
$\{\alpha^{}_1, \dots, \alpha^{}_n\}$,~ $0 < |\alpha^{}_i| < 1$, then
the spectrum of ${\cal A}$ is
\be \label{spectrum}
   {\rm spec}({\cal A}) = \{ \alpha^{\bms{a}} \mid \bm{a} 
                          \in \ZZ_{\ge 0}^n\}
\ee
where $\alpha^{\bms{a}} := \alpha_1^{a_1}\cdot \dots \cdot 
                           \alpha_n^{a_n}$.
{}Furthermore, the multiplicities are those suggested by the notation:
\be
   {\rm mult}(\lambda) = \h \{\bm{a} \mid \alpha^{\bms{a}} 
                       = \lambda \} \, .
\ee
In particular, if $ \alpha_1^{} = \dots = \alpha_n^{} = \alpha$,
then $\lambda = \alpha^{|\bms{a}|}$, 
$|\bm{a}| = \alpha_1^{} + \ldots + \alpha_n^{}$, and
$$\rm{mult}(\lambda) = \left( \begin{array}{c}  
 |\bm{a}| + n - 1 \\ |\bm{a}| \end{array} \right) \, ,$$
a formula that is well-known from the $n$-dimensional 
harmonic oscillator.

{\sc Remark}: In \cite{Jia, Jiang} $A$ is assumed to be diagonalizable
and all the eigenvalues are assumed real. 
The diagonalizability of $Q$ guarantees
that of $Q^*=A$ \cite{BM}. By allowing complex-valued functions to enter the 
picture, it is not necessary to assume that the eigenvalues of $A$ 
are real.
Nor do we wish to impose such an assumption since we do not have
this type of control over the spectrum of $A$.

It is quite easy to find eigenfunctions representing these
eigenvalues in terms of the invariant density $f$.
To do so, we choose an eigenbasis $\{v^{}_1, \ldots , v^{}_n\}$
for $A^t$ in $\CC^n$,
using the fact that $A^t$ is diagonalizable.
Any $k\in \CC^n$ can be written as 
$k = \kappa^{}_1 v^{}_1 + \ldots + \kappa^{}_n v^{}_n$
and the $\kappa^{}_j = \kappa^{}_j(k)$ are the corresponding 
coordinate 
functions (which we allow to be $\CC$-valued). 
Now, fix $\bm{b} \in \ZZ_{\ge 0}^n$, define 
$\kappa^{\bms{b}} = \kappa_1^{b_1}\cdot\ldots\cdot\kappa_n^{b_n}$, 
and let 
$\hat{u}(k) := (\kappa(k))^{\bms{b}} \hat{f}(k)$. Then 
we obtain from (\ref{fourierCntsRef})
\begin{eqnarray} \label{eigenFunctions}
\widehat{{\cal A} u}(k) & = & \widehat{X_{\Omega_Q^{}}}(k) 
                              \hat{u}(A^t k) 
   \; = \; \widehat{X_{\Omega_Q^{}}}(k) (\kappa(A^tk))^{\bms{b}} 
         \hat{f}(A^t k) \nonumber \\
   & = & \alpha^{\bms{b}} \, (\kappa(k))^{\bms{b}} \hat{f}(k)
   \; = \; \alpha^{\bms{b}} \, \hat{u}(k).
\end{eqnarray}

Returning from the Fourier domain, this enables us to write down the
eigenfunctions of $\cal A$. 
If $\{v^{\dagger}_1, \ldots , v^{\dagger}_n\}$
denotes the dual basis 
(i.e.\ $v^{\dagger}_i \cdot v^{}_j = \delta^{}_{ij}$),
we define the directional derivative 
${\rm D}_j := v^{\dagger}_j \cdot \nabla$
and obtain
\begin{prop}
The partial derivatives
\be
    {\rm D}^{\bms{b}}f, \; \bm b \in \ZZ_{\ge 0}^n
\ee
are eigenfunctions of the refinement operator ${\cal A}$,
with eigenvalue $\alpha^{\bms{b}}$.
\end{prop}
Some of these derivatives, calculated for our guiding Example, 
are shown in Figure 2.

% *****************************************
%  Figure 2 goes here  derivatives
% ******************************************

\vspace*{8mm}
\begin{figure}[ht]
\centerline{\epsfysize=65mm \epsfbox{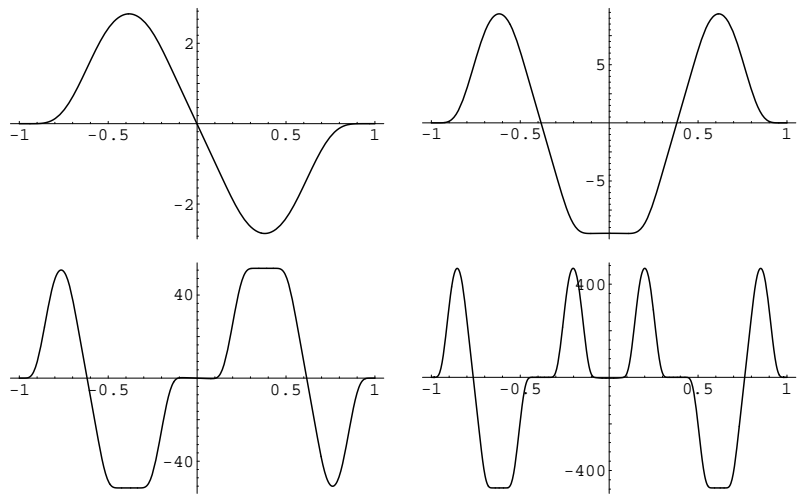}}
\caption{The first 4 derivatives of the invariant density of 
         Example \ref{ex2} for inflation factor $\tau$.}
\end{figure}

\section{Further remarks}

\subsection{Diffraction and a product formula}

The product formula (\ref{prodFormula}) has an interpretation
on the physical side of our picture. To see this, we define
functions $g_s$ and $h_s$ by
\begin{eqnarray} \label{5.2}
   g_s(k) & = & {1\over \h T_s}
       \sum_{v\in T_s} e^{-2\pi i k\cdot v} \\
   h_s(k) & = & {1\over \h\Lambda_s}
       \sum_{w\in\Lambda_s} p(w) e^{-2\pi i k\cdot w} \nonumber
\end{eqnarray}
and, furthermore, functions $g$ and $h$ by
\begin{eqnarray} \label{5.3}
     g(k) & = & \lim_{s\rightarrow\infty} g_s(k) \\
     h(k) & = & \lim_{s\rightarrow\infty} h_s(k) \, . \nonumber
\end{eqnarray}

The existence of $g$ as a well-defined function on $\RR^m$ is a known
consequence of the fact that $T$ is a model set. We now have \cite{BM}:
\begin{prop}
  $\quad h(k) \; = \; \prod\limits_{N=0}^{\infty} g((Q^t)^N k)$.
\end{prop}

The significance of the functions $g$ and $h$ appears in the context
of diffraction. Suppose that $w:L \rightarrow \RR_{\ge 0}$ is some
bounded non-negative function that vanishes off $\Lambda$.
Define the tempered distribution
\be
 \mu_{w} \; = \; \sum_{x \in \Lambda} w(x) \delta_x \nonumber
\ee
where $\delta_x$ is the Dirac measure at $x$. The limit,
as $s \rightarrow \infty$, of the averaged auto-correlation
of this measure, which exists for model sets, is the 
{\em auto-correlation measure} of $\Lambda$
(also called its Patterson function, though it is a distribution)
\be
   \gamma_w \; = \; \lim_{s \to \infty } \frac{1}{\h\Lambda_s}
   \sum_{x,y \in \Lambda_s} w(x)w(y)\delta_{x-y}~.  \nonumber
\ee
Its Fourier transform is a positive measure $\hat{\gamma}_w$  which
is the {\em diffraction pattern} of $\Lambda$. The point
part of this measure is the {\em Bragg spectrum} of $\Lambda$.
In the case that $w$ is the indicator function (i.e constant
value $1$) on a model set $\Lambda$ then $\hat{\gamma}_w$ 
is a pure point measure and we say that $\Lambda$ has a 
{\em pure point spectrum}. In any case, the Bragg
spectrum can be calculated from the simpler {\em function} 
\be
g(\cdot\,;w) \; : \quad g(k;w) 
              \; = \; \lim_{s \to \infty} \frac{1}{\h\Lambda_s}
              \sum_{x \in \Lambda_s} w(x) e^{-2\pi i x\cdot k}
\ee
provided that this limit exits everywhere. In fact \cite{Hof}
\be
 |g(k;w)|^2 \; = \; 
            \hat{\gamma}_w(\{k\}),~ \mbox{for all} ~ k \in \RR^m \, .
\ee

Our functions $g$ and $h$ correspond to the cases when
$w = \cf_\Lambda$ and $w$ is the invariant density $p$, respectively. 
In particular, $h$ allows us an exact description of the 
intensities of the Bragg spectrum of $(\Lambda,p)$. Notice
from the product formula that its support necessarily lies
inside the support of the Bragg spectrum of $\Lambda$ itself.
It goes without saying that both $g$ and $h$ are highly discontinuous
functions which, for a model set, are non-zero only on a
dense point set of zero Lebesgue measure. This set is contained
in the so-called {\em Fourier module} of the model set, i.e.\
the in the set $\pi^{}_1(\tilde{L}^o)$, where 
$\tilde{L}^o = \{ y \in \RR^{m+n} \mid x\cdot y \in \ZZ \;
\mbox{for all } x \in \tilde{L}\}$ is the dual lattice of $\tilde{L}$.

\subsection{Hutchinson measures}

Let $f$ be the invariant density of ${\cal C}(\Omega)$ corresponding
to the compatible similarity $Q$ on $\Lambda = \Lambda(\Omega)$.
There is a corresponding measure $\mu = \mu^{}_f$, with support 
contained in $\Omega$, defined by 
\be
  \mu_f(Y) \; = \; \int_{\RR^n}\cf_Y(x) \mu^{}_f(dx)
           \; = \; \int_{\RR^n}\cf_Y(x) f(x) dx \, .
\ee
We have $\mu^{}_f(\Omega)= 1$. The measure $\mu^{}_f$ 
is invariant in the sense that, if we define $t^*_v \cdot\mu^{}_f$
by  $t^*_v \cdot\mu_f(Y) = \mu_f((t^*_v)^{-1}(Y))$, then
\be \label{invariantdensity}
   \mu^{}_f \; = \; \lim_{s \to \infty} \frac{1}{\h T_s}
\sum_{v\in T_s} t^*_v\cdot\mu^{}_f ~.
\ee

Now fix some $s>0$ and consider the finite set of contractions $t^*_v$
on $\Omega$, where $v \in T_s$. According to \cite{Hutchinson}
there is a unique non-negative Borel measure $\mu_s$ on $\Omega$ 
for which $\mu_s(\Omega) = 1$ and which is {\em invariant} 
in the sense that
\be \label{hutchMeas}
 \mu_s \; = \; \frac{1}{\h T_s}\sum_{v\in T_s} t^*_v\cdot\mu_s ~.
\ee
Furthermore, this measure is the unique fixed point of the process
on the set of regular Borel measures on $\Omega$ with mass 1 
that averages a measure by the right hand side of (\ref{hutchMeas}).
We call this a {\em Hutchinson measure}. Using the Levy continuity
theorem \cite{Bauer} one can show
\begin{prop} The sequence of Hutchinson measures fulfils:
\begin{itemize}
\item[{\rm(i)}] $\hat{\mu}_s(k) = \prod_{N=0}^\infty g_s((A^t)^Nk)$,
for any $s >0$;
\item[{\rm(ii)}] $\{\hat{\mu}_s\} \rightarrow \hat{\mu}^{}_f$, where
the convergence is uniform on compact sets;
\item[{\rm(iii)}]  $\{\mu_s\} \rightarrow \mu^{}_f$, in the 
sense of weak
convergence, i.e.\ $\{\mu_s(\varphi)\} \rightarrow \mu_f(\varphi)$
for all $\varphi\in{\cal C}(\Omega)$.
\end{itemize}
\end{prop}
In particular, for any function $\varphi\in{\cal C}(\Omega)$,
$\mu_s(\varphi)$, for large $s$, is a good estimate of 
$\mu^{}_f(\varphi)$.
Also, starting from {\em any} probability measure on $\Omega$,
the Hutchinson iteration of (\ref{hutchMeas}) will converge to
$\mu_s$ and this procedure, for large $s$, will also give a good
approximation to $\mu^{}_f$.

\subsection{The topology of $L$ and $\Lambda$}

The space of functions ${\cal C}(\Lambda)$ 
on which the averaging operator ${\cal A}$ acts seems
strange. It is defined in terms of the topology
of $\Omega$ in $\RR^n$ and this is very different
from the discrete topology that we see on $\Lambda$ induced
by the topology of its ambient space $\RR^m$.
The appropriate topology for $\Lambda$ (and $L$)
is defined intrinsically as follows \cite{Martin2}. For each compact
set $K$ of $\RR^m$ define
\be
N_K \;=\; N_K(\Lambda)\;:=\;\{v \in L \;|\;  v + (\Lambda \cap K)
= \Lambda \cap (v+K)\}.
\ee
Thus $N_K$ is the set of vectors $v$ for which translation
by $v$ is a bijection of the $K$-patch of $\Lambda$
and onto the $(v+K)$-patch of $\Lambda$.  Note that
the mapping $K \mapsto N_K$ is inclusion reversing.
\begin{prop} {\rm \cite{Martin,Martin2}}
Suppose that $L^* \cap \partial \Omega = \emptyset$. 
Then the collection
of sets $\{N_K \;|\; K \subset \RR^m,~ K ~\mbox{compact} \}$
is a basis of neighbourhoods of $0$ for a topological
group structure on $L$. Furthermore, $\RR^n$ with its standard
topology is the completion of $L$ under the mapping
$()^* :L \rightarrow \RR^n $. With this topology on $L$,
the space ${\cal C}(\Lambda)$
is precisely the space of continuous functions on $L$
whose support lies in $\Lambda$.
\end{prop}
The intuition behind continuity of a function $\phi$ (defined on 
$L$) with respect to this topology
is this: if translation by $v$ is a bijection of two ``large''
patches of $\Lambda$, then $\phi(x+v) - \phi(x)$ is (uniformly) 
small.

\section{Outlook}
The existence of {\em positive} invariant densities naturally
suggests probabilistic interpretations. In \cite{BM2} we study
the spectral properties of certain stochastic sets whose sites 
are selected from that of a model set on a probabilistic basis
according to the density $p$. Effectively, this gives a distribution
of points which, after $*$-mapping them into internal space, 
looks like 
our window shaped by the invariant density $f$. This may provide an
alternative explanation of the recently made observations of such
profiles in real data \cite{Dieter}. Furthermore, as such sets
do have finite entropy density, they might be useful for further
models of entropic stabilization of quasicrystals.

In this article, we have focused on one-component model sets.
It is important for multi-component or multi-coloured systems
to be able to realize the similarity-averaging process and the
existence of invariant densities in a matrix generalization of
what we have done here. This means that we have a finite family
of model sets based on cosets of a common $\ZZ$-module and
matrices of similarity maps between them. In fact this set-up
results in matrix continuous refinement operators and ultimately
again in the existence of invariant densities. What is particularly
interesting in this case is the appearance of a Markov matrix
of weights for the contributions from the various windows relative
to each other. An exposition of this will appear in \cite{BM3}.

\section*{Acknowledgements}
It is our pleasure to thank R.~Q.~Jia, W.~Allegretto, and A.~Hof 
for valuable discussions. 
We thank the Volkswagen Stiftung for support through
the RiP-program at Oberwolfach where this paper was written. 
MB is supported
by the German Science Foundation (DFG) through a 
Heisenberg-fellowship. RVM 
thanks the Tata Institute of Fundamental Research, Bombay and
the Natural Sciences and Engineering Research Council of Canada
for their support in this work.

\end{document}